\documentclass[5p,sort&compress]{elsarticle} % , showpacs, superscriptaddress
\usepackage{bbm, amsmath, amssymb, amsthm, bm, textcomp, nicefrac}
\usepackage[utf8]{inputenc}
\usepackage[T1]{fontenc}
\usepackage[english]{babel}

\usepackage{geometry}
\usepackage{color}

\begin{document}

\title{Linking Measures for Macroscopic Quantum States via Photon-Spin Mapping}

\author[Gen]{F.~Fr\"owis\corref{cor1}}
\ead{florian.froewis@unige.ch}
\author[Gen,Bas]{N.~Sangouard\corref{cor2}}
\author[Gen]{N.~Gisin}
\cortext[cor1]{Corresponding author}
\address[Gen]{Group of Applied Physics, University of Geneva, CH-1211 Geneva 4, Switzerland}
\address[Bas]{Department of Physics, University of Basel, CH-4056 Basel, Switzerland}

\date{\today}

\begin{abstract}
  We review and compare several measures that identify quantum states that are ``macroscopically quantum''. These measures were initially formulated either for photonic systems or spin ensembles. Here, we compare them through a simple model which maps photonic states to spin ensembles.  On the one hand, we reveal problems for some spin measures to handle correctly photonic states that typically are considered to be macroscopically quantum. On the other hand, we find significant similarities between other measures even though they were differently motivated.
\end{abstract}

\maketitle

\section{Introduction}
\label{sec:introduction}

The first experiments that triggered the development of quantum mechanics were conducted by relatively simple means. After the theoretical framework has been established, the effort to experimentally verify some much more demanding predictions like entanglement and nonlocality increased significantly. Nowadays, we master experimental techniques that even led to commercial products such as secure communication and true random number generators. Furthermore, it is by now possible to enter the quantum regime of ``large'' systems; large either in terms of mass, energy or number of involved microscopic constituents. Among other contributions, experimenters brought superconducting devices \cite{andrews_observation_1997,friedman_quantum_2000,hime_solid-state_2006} and massive mechanical oscillators \cite{teufel_sideband_2011,kiesel_cavity_2013} to the quantum regime; one observed interference effects with giant molecules \cite{gerlich_quantum_2011}, entangled diamonds \cite{lee_entangling_2011}, cells \cite{hald_spin_1999}, doped crystals \cite{usmani_heralded_2012} and large spin ensembles \cite{riedel_atom-chip-based_2010,behbood_generation_2014,lucke_detecting_2014}; we also witnessed entanglement between optical modes including hundreds of photons \cite{bruno_displacement_2013,lvovsky_observation_2013}. Arguable, all these experiments show quantum behavior in large systems. But how could one compare them? In which sense is one more macroscopically quantum than another one?
Answers to these and similar questions would allow us to challenge old but still unsolved problems. One of those is the transition between microscopic and macroscopic domain. How and in which sense do large systems become ``classical'' --- after all, they are composed of microscopic particles that are quantum mechanically in nature? Some proposals try to answer suchlike questions within quantum theory. For instance, the decoherence program \cite{zurek_decoherence_2003} provides a mechanism for the loss of quantum correlations that typically becomes stronger for larger systems. Other ideas suggest a solution by extending the theory as it is done, for example, in collapse models \cite{ghirardi_unified_1986,bassi_models_2013}. Clearly, these efforts are important for problems concerning the validity and interpretation of quantum mechanics. As well, they immediately yield a practical aspect, for instance, in view of efforts to realize large-scale quantum computing.

Against this background, it is somehow astonishing that a commonly accepted framework of ``macroscopic quantum physics'' is still lacking. The famous gedanken experiment of Schr\"odinger \cite{schrodinger_gegenwartige_1935} ponders on the existence of large objects in a superposition of two classical, distinct states like a cat being dead \textit{and} alive.
Complementary, Leggett \cite{leggett_macroscopic_1980, leggett_testing_2002} argued that, in large systems, there is difference between an accumulated quantum effect originated on a microscopic scale and a ``true'' quantum effect on a macroscopic scale. While the former is undoubtedly an experimental challenge due to the complexity and the large number of degrees of freedom, only the latter is supposed to provide insight into the aforementioned problems. Based on these and other contributions \cite{dur_effective_2002}, many physicists came up with  measures to quantify how ``macroscopically quantum'' a state is \cite{shimizu_stability_2002,bjork_size_2004,shimizu_detection_2005,korsbakken_measurement-based_2007,marquardt_measuring_2008,lee_quantification_2011,frowis_measures_2012,nimmrichter_macroscopicity_2013,sekatski_size_2014}. 
Such mathematical definitions potentially provide a clear view on macroscopic quantum effects. Furthermore, an established definition is the basis for theoretical conclusions of, for example, the stability of macroscopic quantum states with respect to noise \cite{shimizu_stability_2002,frowis_certifiability_2013} and measurement imperfections \cite{sekatski_how_2014}.

Due to the many proposals on the characterization of macroscopic quantum states, it is clearly necessary to compare those measures in order to understand the similarities and differences. First attempts have been made in Refs.~\cite{frowis_measures_2012}, where several measures suitable for spin measures \cite{shimizu_stability_2002,bjork_size_2004,korsbakken_measurement-based_2007,marquardt_measuring_2008,frowis_measures_2012} have been classified. Another work \cite{volkoff_measurement-_2014} applies some measures \cite{dur_effective_2002,korsbakken_measurement-based_2007,marquardt_measuring_2008,frowis_measures_2012} to a specific multi-mode photon state. The ultimate goal is to provide a general framework for macroscopic quantum effects that covers all important physical systems. With this, one is able to directly compare different systems. For instance, one could then compare experiments on trapped ions with massive objects in the superposition of spatial positions (see additional remarks in Sec.~\ref{sec:discussion}).

In this paper, we aim to continue this research line by bridging measures that were formulated for spin systems \cite{shimizu_stability_2002,bjork_size_2004,korsbakken_measurement-based_2007,marquardt_measuring_2008,frowis_measures_2012} and for single photonic modes \cite{lee_quantification_2011,sekatski_size_2014} (some of them are valid for both systems). To this end, we use a simple model of a photon-spin mapping, in particular, the absorption of a photonic state into a spin ensemble. Under the assumption that the properties of the photonic state are completely mapped to the spin ensemble, we have a tool at hand to analyze and compare in which sense states are macroscopically quantum according to different measures.

In the following, we draw some conclusions for several measures based on this mapping. As we will see later, it is necessary that the mean photon number, $N$, of the considered state is much smaller than the number of the spins in the ensemble, $M$. After the mapping, $N$ corresponds to the excitation level of the spins. In this regime, we observe that some measures for spin states behave differently than in the case where $N$ is comparable with $M$, which is the regime where they have been studied so far. Apparently, through this work we also understand better the present proposals and learn about the implications of their initial intuition.

On the other side, we find that there are tight mathematical connections between certain measures, even though the physical motivation for introducing them is apparently very different. We conclude therefore that, at least partially, there exists already some consensus on the characterization of macroscopic quantum states among the present proposals. 

This paper is structured as follows. In Sec.~\ref{sec:revi-geneva-meas}, we set the basic nomenclature and summarize the existing proposals in the field of macroscopic quantum states. We also review some established implications on the stability. In Sec.~\ref{sec:quantum-memory-map}, we introduce and elaborate on the model for the photon-spin interaction we use to link different measures. Some implications are discussed in Sec.~\ref{sec:impl-macro-meas}. Conclusions and outlook are given in Sec.~\ref{sec:discussion}.

\section{Review of measures for macroscopic quantum states}
\label{sec:revi-geneva-meas}

In this section, we first clarify some subtle but important points for the discussion of macroscopic quantum physics (Sec.~\ref{sec:prel-disc}). Then, in Secs.~\ref{sec:char-via-coarse} and \ref{sec:meas-suit-spin}, we give a rough overview on some measures for macroscopic quantum states that have been proposed so far. In Sec.~\ref{sec:impl-stab-schr}, we discuss some implications on the stability in the presence of noise and imperfections.

\subsection{Prelimiary discussion}
\label{sec:prel-disc}

\textit{Common goal of the measures.---} The common feature of all works that are considered in this paper is to identify among all quantum states those that are \textit{macroscopically quantum}. This is done by defining a function $f(\psi) \geq 0$ (some proposals are even defined for mixed states). The larger $f(\psi)$, the more macroscopically quantum $| \psi \rangle $ is. Often, $f$ is called the \textit{effective size} of $| \psi \rangle $. The \textit{qualitative} distinction between macroscopic and non-macroscopic quantum states based on $f$ is to some extent arbitrary.

\textit{System size.---} All published proposals agree that a quantum state can only be macroscopically quantum if the respective system size is in some sense ``large''. For systems composed of microscopic particles, it is necessary to have many constituents. If one considers one (or few) bosonic modes, we require to have high excitation numbers or high masses. The exact values for having a ``large'' system are not crucial for the present discussion. As we are concerned with spin and photonic systems in this paper, the \textit{system size} is defined as the number of spin-$1/2$ particles, $M$, or the mean photon number, $N$, respectively. 

\textit{Schr\"odinger-cat state vs. macroscopic quantum state.--- } A first distinction of the current literature can be made by the basic form of the states considered to be macroscopic. Some authors \cite{bjork_size_2004,korsbakken_measurement-based_2007,marquardt_measuring_2008,sekatski_size_2014} consider superpositions of two (or a few) ``classical'' states like
\begin{equation}
\label{eq:1}
\left| \psi \right\rangle \propto \left| \psi_0 \right\rangle + \left| \psi_1 \right\rangle .
\end{equation}
In a simplified way, one can say that one seeks a mathematical definition for the verbal characterization that the states $| \psi_0 \rangle $ and $| \psi_1 \rangle $ are ``macroscopically distinct'' (Leggett \cite{leggett_macroscopic_1980}). In the remainder of this paper, we call macroscopic superpositions of the form (\ref{eq:1}) a \textit{Schr\"odinger-cat state}.

On the other side, some proposals \cite{shimizu_stability_2002,shimizu_detection_2005,lee_quantification_2011,frowis_measures_2012,nimmrichter_macroscopicity_2013} do not require a specific form of the quantum state and may even allow for mixed states. States that are macroscopically quantum due to these definitions are here called \textit{macroscopic quantum states} (see examples below). 
If any confusion is excluded, we also use this term as an umbrella term that includes the concept of a Schr\"odinger-cat state.

\textit{Scaling versus absolute numbers.---} Experiments that aim to generate macroscopic quantum states may be interested in states $| \psi \rangle $ with large $f(\psi)$. In such a case, the absolute value of the normalized $f(\psi)$ is the figure of merit. In turn, in many theoretical studies, one does not only consider a fixed system size but more abstractly \textit{state families}. A state family is a prescription that assigns to any system size a quantum state $| \psi \rangle $. Then, one can investigate, for example, the scaling of $f(\psi)$ with the system size.\footnote{The scaling of a function $f$ with $N$ is defined as $f = O(g(N)) :\Leftrightarrow \lim_{N\rightarrow \infty} f/g$ exists and is strictly positive.} The larger the scaling for $f(\psi)$, the more macroscopically quantum is the respective state family $| \psi \rangle $. In this paper, we focus on how the considered measures scale. In particular, we say that $| \psi \rangle $ is macroscopically quantum with respect to $f(\psi)$ if the respective state family scales linearly with the system size. Furthermore, two measures are said to be compatible if they identify the same class of macroscopic quantum states.

\textit{Subjectivity.---} Finally, we add a comment that was, in similar words, already made in several other works. In order to identify macroscopic quantum states, one has to introduce some kind of restrictions or some ``structure'' within the Hilbert space. This is similar to entanglement theory, where one has to define a tensor structure in a high-dimensional Hilbert space to distinguish between separable and entangled states.

The restrictions for the measures of macroscopic quantum states are to some extent subjective. This fact could be used to criticize the measures. However, in many proposals, the predefined structure is motivated by actual limitations we encounter in nature (and therefore in the lab). For example, the restriction to certain (local) measurements \cite{shimizu_stability_2002,korsbakken_measurement-based_2007,sekatski_size_2014}, local interactions \cite{frowis_measures_2012} or certain noise models \cite{lee_quantification_2011} reflect such strong limitations. We believe that they are necessary in order to identify a qualitative difference between macroscopic and accumulated microscopic quantum effects.

Note that, for spin systems, a measurement or interaction is called \textit{local}, if the respective operator can be written as a sum of terms where each term has nontrivial support only on one particle. For example, the interaction with an external magnetic field oriented in $z$ direction is generated by 
\begin{equation}
\label{eq:9}
J_z = \sum_{i=1}^M \sigma_z^{(i)},
\end{equation} 
where $\sigma_z^{(i)} \equiv \mathit{id}_{\mathbbm{C}^2}^{\otimes i-1} \otimes \sigma_z \otimes  \mathit{id}_{\mathbbm{C}^2}^{\otimes M-i}$ is a Pauli operator acting on the $i$th spin. For the sake of simplicity, we only consider operators that are symmetric with respect to qubit permutations. Furthermore, all local operator considered in this paper are normalized to a spectral radius that equals the number of qubits.\footnote{This is also the reason why we omit the factor $1/2$ in Eq.~(\ref{eq:9}).} We note that many of the following concepts can be generalized to ``quasi-local'' operators in the sense that each addend has nontrivial support on $O(1)$ qubits (e.g., nearest-neighbor interaction).

\subsection{Measures for photonic systems}
\label{sec:char-via-coarse}\label{sec:meas-wign-funct}

We now review two measures that are naturally defined for photonic systems.

\textit{Coarse-grained-based measure \cite{sekatski_size_2014}.--- }
Consider two states $| \psi_0 \rangle $ and $| \psi_1 \rangle $. Assume there exists a measurement that distinguishes them with high success probability. To call the two states macroscopically distinct, in Ref.~\cite{sekatski_size_2014} it is required that they are distinguishable even with very coarse-grained (i.e., ``classical'') detectors.

The specific definition is based on photon-number detection in one-mode states. The coarse-graining is introduced via a Gaussian distribution that fixes the probability to measure $m$ photons when indeed there have been $n$ present, $p(m|n) = 1/(\sqrt{2 \pi} \sigma) \exp (-(n-m)^2/(2\sigma^2))$, where $\sigma \geq 0$ is the parameter that adjusts the measurement accuracy. The effective size of superposition (\ref{eq:1}) is then the maximal value $\sigma$ such that $| \psi_0 \rangle$ and $| \psi_1 \rangle$ are distinguishable with a success probability $P_s$ of at least $P_g \in (1/2,1]$, which is free parameter of the measure. Thus, the effective size is reads
\begin{equation}
\label{eq:25}
\mathrm{Size}_{P_g} = \max_{\sigma: P_s \geq P_g}2 \sqrt{2} \mathrm{erf}^{-1}(P_g - 1)\sigma\end{equation}
for general $P_g$, where $\mathrm{erf}$ is the error function. As an example, one finds that $\mathrm{Size}_{2/3} \approx \max_{\sigma: P_s \geq 2/3} 0.86 \sigma$.  

Let us examine the measure for the photonic state
\begin{equation}
\label{eq:34}
\left| \psi \right\rangle = \frac{1}{\sqrt{2}}\left( \left| 0  \right\rangle +\left|2 N \right\rangle   \right),
\end{equation} which is often considered to be a Schr\"odinger-cat state. It is straightforward to see that $\sigma$ can be up to $O(N)$ for any value $P_g\in (1/2,1)$; hence it identifies $| \psi \rangle $ as macroscopically quantum.

As a second example, consider the archetypal macroscopic photonic superposition  
\begin{equation}
\label{eq:2}
\left| \Psi_{\alpha} \right\rangle = \frac{1}{\sqrt{2(1+\exp(-2 |\alpha|^2))}}\left( \left| \alpha \right\rangle + \left| -\alpha \right\rangle  \right),
\end{equation}
where $| \pm\alpha \rangle = \mathcal{D}_{\pm \alpha} \left| 0 \right\rangle $ is a coherent state. The operator $\mathcal{D}_{\alpha} = \exp(\alpha a^{\dagger} - \alpha^{*} a)$ (with $a$ as the annihilation operator and $\alpha \in \mathbbm{C}$) is the so-called displacement operator and $| 0 \rangle $ is the vacuum state. The states $| \alpha \rangle $ and $| -\alpha \rangle $ are not distinguishable with photon number measurements. 
However, its macroscopic quantum nature can be revealed by extending the definition of the measure to all ``state of the art'' measurements, like homodyne measurements. This modification enlarges the class of Schr\"odinger-cat states to states like $| \Psi_{\alpha} \rangle $.

We emphasize that this measure is not invariant under displacement operations. This can be illustrated by a third example, which is the displacement of a single photon in the superposition of two different modes \cite{sekatski_proposal_2012}, 
\begin{equation}
\label{eq:37}
\left| \psi^{-}_{\alpha} \right\rangle = \frac{1}{\sqrt{2}}
  \mathcal{D}_{\alpha}\otimes id \left( \left| 0,1 \right\rangle -
    \left| 1,0 \right\rangle \right).
\end{equation}
In order to apply the coarse-grained-based measure, consider a basis change to $\left| \pm \right\rangle \propto \left| 0 \right\rangle \pm \left| 1 \right\rangle $. The question is thus how well can we distinguish between $\mathcal{D}_{\alpha}\left| + \right\rangle $ and $\mathcal{D}_{\alpha}\left| - \right\rangle $. As shown in Ref.~\cite{sekatski_size_2014}, the effective size reads 
\begin{equation}
\label{eq:38}
\mathrm{Size}_{P_g}(\left| \psi^{-}_{\alpha} \right\rangle ) = 2 \alpha \mathrm{erf}^{-1}(2P_g -1) \sqrt{\frac{4}{\pi(2P_g-1)^2}-2},
\end{equation}
that is, it scales with the square root of the photon number in the regime where $P_g \in (1/2, 1/2(1+\sqrt{2/\pi}))\approx (0.5,0.8989)$.

Besides the obvious intuition to define macroscopic distinctness by realistic detectors, there is a more abstract motivation to connect this measurement to the notion of Schr\"odinger-cat states. The specific noise model is such that nearby eigenstates of the photon number operator are less distinguishable. The distinctness of two states that are hardly distinguishable in the presence of noise is hence based on ``microscopic details''. The noise model helps to reveal how much their distinguishability depends on microscopic and macroscopic differences, respectively.

\textit{Counting oscillations in the Wigner function~\cite{lee_quantification_2011}.--- }
The motivation for this measure is most easily gathered by the following example. Compare the superposition $| \Psi_{\alpha} \rangle $ with the incoherent mixture $1/2\left|\alpha  \right\rangle\!\left\langle \alpha\right| + 1/2\left| -\alpha \right\rangle\!\left\langle -\alpha\right| $. The inspection of the respective Wigner functions shows that both states exhibit strong peaks around the values $\pm \alpha$. But while for the incoherent mixture one has a sum of two (positive) Gaussian functions, the Wigner function of $| \Psi_{\alpha} \rangle $ shows strongly fluctuating oscillations between the peaks.

The authors of Ref.~\cite{lee_quantification_2011} systematized this insight by defining the effective size of a state $\rho$ that measures the frequency and the amplitude of the Wigner function. They find that this quantity can be calculated without directly invoking the Wigner function. One has 
\begin{equation}
\label{eq:4}
\mathcal{I}(\rho) = - \mathrm{Tr}[\rho \mathcal{L}(\rho)] + \mathrm{Tr}\rho^2/2
\end{equation}
with 
\begin{equation}
\label{eq:5}
\mathcal{L}(\rho) = -\sum_{m} \left[ a_m \rho a^{\dagger}_m - \frac{1}{2} \rho a^{\dagger}_m a_m - \frac{1}{2} a^{\dagger}_m a_m \rho \right],
\end{equation}
where $m$ is the index for the different modes of $\rho$. Note that the last term was proposed later \cite{jeong_reply_2011} in order to guarantee positivity of the measure. For pure, single-mode states, $| \psi \rangle $,Eq.~(\ref{eq:4}) reduces to 
\begin{equation}
\label{eq:6}
\mathcal{I}(\psi) = \langle a^{\dagger} a \rangle_{\psi} - \left| \langle a \rangle_{\psi} \right|^2 + \frac{1}{2}.
\end{equation}

One easily sees that the Fock state $| N \rangle $ thus is a macroscopic quantum state as $\mathcal{I} = N+1/2$. The superposition $| 0 \rangle + \left| 2N \right\rangle $ exhibits the same effective size and is hence not more macroscopically quantum as $| N \rangle $.

For $| \Psi_{\alpha} \rangle $, one has that $\mathcal{I} = \langle a^{\dagger} a \rangle + 1/2= \left| \alpha \right|^2 \tanh \left| \alpha \right|^2+1/2$, while the incoherent sum is zero-valued.

Note that $\mathcal{I}(\rho)$ is invariant under displacing $\rho$, that is, $\mathcal{I}(\mathcal{D}_{\alpha}\rho \mathcal{D}_{\alpha}^{\dagger}) = \mathcal{I}(\rho)$. Therefore, the state $| \psi^{-}_{\alpha} \rangle $ is not considered to be macroscopically quantum by this measure.

\subsection{Measures suitable for spin systems}
\label{sec:meas-suit-spin}

In this subsection we mention the measures that are suitable for spin systems (even if they are not exclusively defined for spins). This part is compactly written since a more extended review is already given in either Refs.~\cite{frowis_measures_2012,volkoff_measurement-_2014}. As a simple example for the measures, we consider the Greenberger-Horne-Zeilinger (GHZ) state 
\begin{equation}
\label{eq:3}
\left| \mathrm{GHZ} \right\rangle = \frac{1}{\sqrt{2}} \left( \left| 0 \right\rangle ^{\otimes M} + \left| 1 \right\rangle ^{\otimes M} \right),
\end{equation}
with $| 0 \rangle $ and $| 1 \rangle $ as the eigenstates of $\sigma_z$ with eigenvalue $1$ and $-1$, respectively.

\textit{Index $p$ \cite{shimizu_stability_2002,shimizu_detection_2005}.--- } Consider a system of $M$ components and a local measurement $A$ (e.g., $J_z$). The motivation for introducing the \textit{index} $p$ is the observation that a typical classical probability distribution exhibits a standard deviation in the order of $\sqrt{M}$. The ratio of standard deviation and spectral radius vanishes in the limit of large $N$. However, there exists quantum states and corresponding local observables where the standard deviation is also in the order of $M$, meaning that even for very large system sizes, one observes persistent fluctuations in the measurement statistics.

The index $p$ is therefore defined as 
\begin{equation}
\label{eq:7}
\max_{A: \mathrm{local}} \mathcal{V}_{\psi}(A) = O(M^p),
\end{equation}
where $\mathcal{V}_\psi(A) = \langle A^2 \rangle_{\psi} - \langle A \rangle_{\psi}^2$ is the variance of $A$ with respect to $| \psi \rangle $. The maximization is carried out over the set of symmetric, normalized local operators. If $p = 2$, the $| \psi \rangle $ is called a macroscopic quantum state (in Ref.~\cite{shimizu_stability_2002}, these states are called \textit{anomalously fluctuating states}). The GHZ state yields $\mathcal{V}( J_z) = M^2$, which is the maximal value.

The extension to mixed states is nontrivial. An incoherent mixture $\left| 0 \right\rangle\!\left\langle 0\right| ^{\otimes M} + \left| 1 \right\rangle\!\left\langle 1\right| ^{\otimes M}$ leads to the same variance, but is typically not considered as a macroscopic quantum state. In Ref.~\cite{morimae_superposition_2010} (see also Ref.~\cite{shimizu_detection_2005}), the so-called \textit{index} $q$ is introduced as
\begin{equation}
\label{eq:8}
\max(M, \max_{A:local}\lVert \left[ A,\left[A,\rho \right] \right] \rVert_1) = O(M^q)
\end{equation}
where $\lVert X \rVert_1 = \mathrm{Tr} \sqrt{X X^{\dagger}} $ is the trace norm of $X$. A tight connection to the index $p$ was shown in the case of pure $\rho$.

\textit{Measure based on ``increasing the interferometric utility'' \cite{bjork_size_2004}.--- } Another way to characterize Schr\"odinger-cat states is to ask how much more useful is superposition (\ref{eq:1}) compared to its components $| \psi_0 \rangle $ and $| \psi_1 \rangle$. The choice of the specific task is clearly a crucial aspect of this kind of characterization. Nevertheless, there does not seem a conceptual difference between choosing a certain task and restricting oneself to certain (local) Hamiltonians and measurements (as done for other measures). The authors of Ref.~\cite{bjork_size_2004} focus on interferometry experiments with a given generator $A$ under idealized conditions and introduce the quantity $\mathcal{M} = \theta_{\mathrm{sing}}/\theta_{\mathrm{sup}}$, where $\theta_{\mathrm{sing}}$ and $\theta_{\mathrm{sup}}$ are the sensitivities of the single states $| \psi_i \rangle $ $ (i = 0,1)$ and the superposed state $| \psi \rangle $, respectively. In principle, this idea is applicable to photonic and spin systems.

In Ref.~\cite{frowis_measures_2012}, it is argued that one should square $\mathcal{M}$ in order to compare this measure to others like \cite{korsbakken_measurement-based_2007,marquardt_measuring_2008,frowis_measures_2012}. As shown in Ref.~\cite{bjork_size_2004}, one then finds
\begin{equation}
\label{eq:10}
\mathcal{M}^2 \approx \frac{(\langle A \rangle_{\psi_1} - \langle A \rangle_{\psi_0})^2}{\mathcal{V}_{\psi_0}(A) +  \mathcal{V}_{\psi_1}(A)}. 
\end{equation}

The quantity $\mathcal{M}$ is defined only when $| \psi_i \rangle $ are not eigenstates of $A$. The later happens, for instance, to the GHZ state (\ref{eq:3}) with the generator $A = J_z$. This issue can be fixed by specifying the \textit{maximum} benefit of $| \psi \rangle $ compared to \textit{maximum} benefit of the single components, where the maximum is taken over the all ``feasible'' generators (where the term feasible is clearly to be defined). The maximal utility for $| 0 \rangle ^{\otimes N}$ is given --in the case of the restriction to local operators-- for $A = J_x \mathrel{\mathop:}=  \sum_i \sigma_x^{(i)}$. Then, we find for the GHZ state $\mathcal{M}^2 = 2 M$.

\textit{Measurement-based measure \cite{korsbakken_measurement-based_2007}.--- } The analysis of Schr\"odinger's gedanken experiment reveals that a macroscopic superposition is composed of two states that are different in a macroscopically large number of ``sites''. Taking the cat example literally, the dead and the alive cat differ in every biological cell. Translated to a mathematical definition, the authors of Ref.~\cite{korsbakken_measurement-based_2007} call the many-particle superposition (\ref{eq:1}) macroscopically quantum if one can find a macroscopic number of distinct groups such that the measurement of one such group let one distinguish between $| \psi_0 \rangle $ and $| \psi_1 \rangle $. For a more flexible characterization, one introduces a minimal success probability $P_{\mathrm{min}} = 1- \delta$, which has to be overcome by each group. The one-shot probability for a successful distinction is given by the trace norm of the reduced density operators. Define $\rho_i^{(k)} = \mathrm{Tr}_{M \backslash k} \left| \psi_i \right\rangle\!\left\langle \psi_i\right| $, where $k$ is a subset of all qubits. The success probability then equals 
\begin{equation}
\label{eq:11}
P_{\mathrm{S}}^{(k)} = \frac{1}{2} + \frac{1}{4} \lVert \rho_0^{(k)} - \rho_1^{(k)} \rVert_1.
\end{equation}
The effective size is defined as 
\begin{equation}
\label{eq:12}
C_{\delta}(\psi) = M/ n_{\mathrm{min}},
\end{equation}
where $n_{\mathrm{min}}$ is the minimal number of qubits in order to distinguish between $| \psi_0 \rangle $ and $| \psi_1 \rangle $ with $P_{\mathrm{S}}^{(n_{\mathrm{min}})} \geq 1-\delta$.

As an instance, consider the GHZ state. The measurement of each qubit in the basis $| 0 \rangle $ and $| 1 \rangle $ gives perfect distinguishability. Therefore, $n_{\mathrm{min}} = 1$ and $C_1(\mathrm{GHZ}) = M$.

\textit{Measure based on the notion of a ``microscopic step'' \cite{marquardt_measuring_2008}.--- } Another way to characterize macroscopically distinct states is to introduce the concept of a \textit{microscopic step}. In the case of qubits, the natural unit is the creation and annihilation of a spin excitation, that is, $\sigma_x \left| 0 \right\rangle  = \left| 1 \right\rangle$. Then, one ``counts'' how many of those steps one has to apply to, say, $| \psi_0 \rangle $ \textit{on average} in order to reach $| \psi_1 \rangle $. The larger this number, the more macroscopically distinct are these two states. For example, the two components of the GHZ state are macroscopically far apart, since one has to apply qubit flip operations to all qubits of $| 0 \rangle ^{\otimes M}$ to reach $| 1 \rangle ^{\otimes M}$.

To make this idea mathematically more rigorous, start with the subspace $\mathcal{H}_0 = \mathrm{span}\left\{ \left| \psi_0 \right\rangle  \right\}$. For $d>0$, iteratively construct $\tilde{\mathcal{H}}_d$ as the span of all states that can be generated by applying local operators like $\sigma_x^{(i)}$ to all states $| \phi \rangle \in \mathcal{H}_{d-1}$ and define $\mathcal{H}_d = \tilde{\mathcal{H}}_d \backslash \mathcal{H}_{d-1}$. In words, $\mathcal{H}_d$ contains all states that are one (nontrivial) microscopic step way from $\mathcal{H}_{d-1}$ (and also from $\mathcal{H}_{d+1}$). Then, represent $| \psi_1 \rangle $ in terms of this Hilbert space structure, that is, $| \psi_1 \rangle = \sum_{d = 0}^{\infty} \lambda_d \left| v_d \right\rangle $ with $| v_d \rangle \in \mathcal{H}_d$. The effective size $\bar{D}$ is defined as average distance of $\left| \psi_1 \right\rangle $ to $| \psi_0 \rangle $ via
\begin{equation}
\label{eq:13}
\bar{D} = \sum_{d=0}^{\infty} \left| \lambda_d \right|^2 d.
\end{equation}
For $| \psi_0 \rangle = \left| 0 \right\rangle ^{\otimes M}$, one easily sees that $| 1 \rangle ^{\otimes M} \in \mathcal{H}_N$. Therefore, $\bar{D} = M$, which coincides with the intuition.

While this definition is in general not limited to qubit systems, one might disagree on the right microscopic step in other systems. For example, the annihilation operator $a$ may be a good candidate in one-mode photonic systems, but this is not clear at all.

\textit{Measure based on quantum Fisher information \cite{frowis_measures_2012}.--- } The goal of this proposal is to identify a macroscopic quantum effect that is clearly distinct from any accumulated microscopic quantum effect. To this end, consider the unitary time evolution and ask, for given initial state and system Hamiltonian $H$, what is the minimal time in order that the evolved state becomes orthogonal to the initial state. This time is called \textit{orthogonalization time} $\tau$. It turns out that, for pure states $| \psi \rangle $, the variance of $H$ gives a lower bound on $\tau$, that is, $\tau \geq \pi/ (2 \sqrt{\mathcal{V}_{\psi}(H)})$ \cite{mandelstam_uncertainty_1945}.  As in the discussion of the index $p$, one observes that, for local Hamiltonians, some entangled states like the GHZ state show a different scaling of the variance than separable states. It is therefore possible to reach much lower values of $\tau$ by using entangled states.

Concerning mixed states, one notices that an improvement of the bound on $\tau$ is by using the so-called quantum Fisher information (QFI) $\mathcal{F}_{\rho}(H)$; see, for example, Ref.~\cite{braunstein_statistical_1994} for the precise mathematical definition. Here, we just note that for pure states, one has $\mathcal{F}_{\psi}(H)  = 4 \mathcal{V}_{\psi}(H)$ and in general $\mathcal{F}_{\rho}(H)  \leq 4 \mathcal{V}_{\rho}(H)$. One can show \cite{frowis_kind_2012} that $\tau \geq \pi/ \sqrt{\mathcal{F}_{\rho}(H)}$.

The definition of macroscopic quantum states based on the QFI is now the following. Given a state $\rho$, one maximizes $\mathcal{F}_{\rho}(H)$ within the set of all ``feasible'' Hamiltonians. In Ref.~\cite{frowis_measures_2012}, the set of local operators is chosen. The effective size is then determined as 
\begin{equation}
\label{eq:14}
\mathcal{N}_{\mathrm{eff}}(\rho) = \max_{H: \mathrm{local}} \frac{\mathcal{F}_{\rho}(H)}{4 M},
\end{equation}
where the normalization ensures that $\max_{\rho} \mathcal{N}_{\mathrm{eff}}(\rho) = M$. Again, the maximum is attained by the GHZ state. The interpretation of $\mathcal{N}_{\mathrm{eff}}$ is the following. A state that exhibits a large QFI gives rise to more rapid changes in the expectation values of some observables, for example, the projection onto the initial state. Its experimental verification is a macroscopic quantum effect, since it cannot be reproduced by separable states nor by states that exhibit only few-particle entanglement \cite{frowis_kind_2012,toth_multipartite_2012,hyllus_fisher_2012}. 

Based on Eq.~(\ref{eq:14}), one can also formulate a measure for Schr\"odinger-cat states (\ref{eq:1}) by defining  
\begin{equation}
\label{eq:15}
\mathcal{N}_{\mathrm{eff}}^{\mathrm{rF}}(\psi) = \frac{\mathcal{N}_{\mathrm{eff}}(\psi)}{\frac{1}{2}\mathcal{N}_{\mathrm{eff}}(\psi_0) + \frac{1}{2}\mathcal{N}_{\mathrm{eff}}(\psi_1)}.
\end{equation}
This quantity is called \textit{relative Fisher information}.

\textit{Classification.--- } The measures that are presented in Sec.~\ref{sec:meas-suit-spin} were classified in Ref.~\cite{frowis_measures_2012}. Under certain restrictions, which are necessary for some measures to be well-defined (see \ref{sec:restr-relat-} and Ref.~\cite{frowis_measures_2012}), one finds the general result that 
\begin{equation}
\label{eq:16}
\begin{split}
  \mathcal{M}^2=  O(M) &\Leftrightarrow
  \mathcal{N}^{\mathrm{rF}}_{\mathrm{eff}} =  O(M) \Rightarrow \\C_{\delta} =
  O(M) &\Leftrightarrow \bar{D} = O(M) \Rightarrow \\p = 2
  &\Leftrightarrow \mathcal{N}_{\mathrm{eff}} = O(M).
\end{split}
\end{equation}
Relations (\ref{eq:16}) should be read in the following way. Suppose, for a given quantum states $| \psi  \rangle $ as in Eq.~(\ref{eq:1}), one finds that $\mathcal{M}^2 = O(M)$. This holds if and only if $\mathcal{N}_{\mathrm{eff}}^{\mathrm{rF}} = O(M)$ and it implies that, for example, $C_{\delta} = O(M)$. The first strict implication (from the first to the second line in Eq.~(\ref{eq:16})) distinguishes between two different concepts within the set of Schr\"odinger-cat states. The implication from the second to the third line in the same relation shows that the concept of macroscopic quantum state is strictly more general and includes Schr\"odinger-cat states defined via the proposals \cite{bjork_size_2004,korsbakken_measurement-based_2007,marquardt_measuring_2008}. Examples of such a macroscopic quantum state that are not a Schr\"odinger-cat state (according to Refs.~\cite{bjork_size_2004,korsbakken_measurement-based_2007,marquardt_measuring_2008}) are states generated by optimal covariant cloner \cite{frowis_are_2012}, or, more generally, certain spin-squeezed states.

\subsection{Implications on stability of Schr\"odinger-cat states}
\label{sec:impl-stab-schr}

While archetypal Schr\"odinger-cat states like the GHZ state has been identified as important quantum states in large systems, it also became more and more evident that there seems to be a tight connection between Schr\"odinger-cat states and their fragility under noise. This intuition was recently supported by mathematical statements \cite{shimizu_stability_2002,frowis_certifiability_2013,sekatski_how_2014}.

Macroscopic quantum states in the sense that $p = 2$ were shown to be fragile with respect to certain \textit{correlated} noise that is generated by local operators \cite{shimizu_stability_2002}. With Eq.~(\ref{eq:16}), this implies instability also with respect to all other measures presented in Sec.~\ref{sec:meas-suit-spin}. On the contrary, states that exhibit $p = 1$ are stable in the sense that there does not exist any such noise model where the total decoherence rates exceeds the sum of the individual noise rates \cite{shimizu_stability_2002}. Note that there are similarities to the statements of Ref.~\cite{lee_quantification_2011} concerning the stability of macroscopic states according to Eq.~(\ref{eq:4}).

In Ref.~\cite{frowis_certifiability_2013}, the effect of \textit{uncorrelated} noise models on Schr\"odinger-cat states are investigated. Consider the set of states (\ref{eq:1}) whose effective sizes according to Refs.~\cite{korsbakken_measurement-based_2007} scale as $O(M)$. One can show that if a state from this set is subject to specific local noise channels (like local depolarization noise), the success probability to distinguish this noisy state from initially orthogonal states decreases exponentially in $M$. In particular, one cannot efficiently distinguish $| \psi_0 \rangle + \left| \psi_1 \right\rangle $ from the mixture $| \psi_0 \rangle\!\langle \psi_0| + \left| \psi_1 \right\rangle\!\left\langle \psi_1\right|  $. This implies that one cannot efficiently verify the experimental generation of Schr\"odinger-cat states in a strict sense.\footnote{See Ref.~\cite{frowis_certifiability_2013} for details how to modify the Schr\"odinger-cat states to avoid this kind of instability. For the prize of reducing the effective size, one can gain stability with respect to this criterion.} Note that this result was --so far-- not shown for more general macroscopic quantum states identified by the index $p$ and the QFI.

Questions regarding the connection of Schr\"odinger-cat states and measurement imperfections are considered in Ref.~\cite{sekatski_how_2014}. There, it is shown that the more distinguishable two one-mode photonic states are with respect to coarse-grained measurements (see Sec.~\ref{sec:char-via-coarse}), the more fragile is the detection of the relative phase of their superposition. Measurement imperfections such as insensitivity and coarse-graining renders an experimental verification of the Schr\"odinger-cat state more challenging as the effective size of the state increases.

The results of Refs.~\cite{frowis_certifiability_2013,sekatski_how_2014} show a certain duality of Schr\"odinger-cat states. The more two states are macroscopically distinct, the harder it is to observe the relative phase between the respective superposition. The loss of phase information, however, reduces the coherent superposition to an incoherent mixture, which is in general not macroscopically quantum anymore.

\section{Mapping between photonic states and spin ensembles}
\label{sec:quantum-memory-map}

In this section, we are going to introduce and derive the main tool for the comparison between measures for photonic and spin systems, respectively. It is a mapping of photonic states to spin states. The spins represent two-level atoms, which are initially in the ground state. They fully absorb the incoming light field and are thus excited. This allows us to investigate the effective size for photonic states with measures that are \textit{a priori} not formulated for this system. 

The total Hilbert space is the tensor product of a photonic field and $M$ two-level systems, that is, $\mathcal{H} =\mathcal{H}_{\mathrm{ph}}\otimes \mathcal{H}_{\mathrm{s}} \equiv \ell^2(\mathbbm{C}) \otimes \mathbbm{C}^{2 \otimes M}$. The ground state and the excited state of a single spin are denoted as $| g \rangle $ and $| e \rangle $, respectively. We model the Hamiltonian as 
\begin{equation}
\label{eq:17}
H = \chi (a \otimes J_{+} + a^{\dagger} \otimes J_{-}),
\end{equation}
where $\chi \in \mathbbm{R}$ is the coupling strength, $J_{+} = \sum_{i=1}^M \left| e \right\rangle\!\left\langle g\right| ^{(i)}$ and $J_{-} = J_{+}^{\dagger}$. The operators $J_{+}$ and $J_{-}$ create and annihilate atomic excitations, respectively, which are symmetrically distributed within the atomic ensemble. This Hamiltonian does not take certain details into account such as the spatial distribution of the spins and the incoming photons, or any propagation effects. Nevertheless, it is in first order a good approximation to the absorption step in a quantum memory \cite{hammerer_quantum_2010}.

It is convenient to introduce the so-called Dicke states 
\begin{equation}
\label{eq:18}
\left| M,k \right\rangle \propto \sum_{\pi_l} \pi_l \left| e \right\rangle ^{\otimes k} \otimes \left| g \right\rangle ^{\otimes N-k},
\end{equation}
where one sums over all particle permutations $\pi_l$. The states $| M,k \rangle $ are properly normalized and form a basis in the subspace of permutation symmetric states. The operators $J_{\pm}$ together with $J_z = \left[ J_{+},J_{-} \right]$ generate a so-called $SU(2)$ algebra with $\left[ J_z, J_{\pm} \right] = \pm 2J_{\pm}$. One can easily show that $J_z \left| M,k \right\rangle = (-M + 2k)\left| M,k \right\rangle$ and $J_{\pm} \left| M,k \right\rangle = C_{\pm} \left| M,k \pm 1 \right\rangle $ with $C_{\pm}^2 = M/2(M/2+1) - (M/2-k)(M/2-k \mp 1)$. With this, it is easy to see that, e.g., $H \left| 1 \right\rangle \otimes \left| M,0 \right\rangle = \chi \sqrt{N} \left| 0 \right\rangle \otimes \left| M,1 \right\rangle $.

We now derive an approximate formula for the propagator $U = \exp(-i H t)$, which becomes exact in the limit of infinite number of spins. We remark that dealing with large spin ensembles is necessary in order to ensure that the mapping from the photonic mode to the spins preserves all properties of the light mode. 
Another important point is that the final results are independent of $M$ for $M\gg N$, which is \textit{a posteriori} verified, see Sec.~\ref{sec:impl-macro-meas}. For the derivation, we first consider the operators $X_+$ and $X_- = X^{\dagger}_{+}$ and $X_3 \mathrel{\mathop:}=  [X_+,X_-]$. If the relation $\left[ X_{3}, X_{\pm} \right] =\pm 2c^2 X_{\pm}, c \in \mathbbm{R}$, is fulfilled, the three operators generate a $SU(2)$ algebra. Then, one can show for $V = e^{\lambda (X_+ + X_{-})}$ that
\begin{equation}
\label{eq:19}
V = e^{\tanh(c \lambda)/c X_{-} } \cosh(c \lambda)^{X_3/c^2} e^{\tanh(c \lambda)/c X_{+} }.
\end{equation}
This is done by the ansatz $V = \exp(f(\lambda) X_{-}) $ $ \exp(g(\lambda) X_3) $ $\exp(h(\lambda) X_{+})$. Differentiating $V$ with respect to $\lambda$ and successive application of the identity 
\begin{equation}
\label{eq:20}
e^A B e^{-A} = \sum_{n= 0}^{\infty} \frac{1}{n!} \underbrace{[ A,\dots[A }_{n \text{ times}},B]\dots],
\end{equation}
the arising differential equations are solved with the boundary conditions $f(0) = g(0) = h(0) = 0$.

Now, the problem is that $X_{+} = a^{\dagger} \otimes J_{-}$ does not give rise to a $SU(2)$ algebra, because of this assignment it follows that $X_3 = - a^{\dagger} a \otimes J_z - J_{+}J_{-}$ does not fulfill the necessary commutation relation with $X_{\pm}$. However, if we consider only states $| \Psi \rangle \in \mathcal{H}$ with a comparatively low excitation number $\langle a^{\dagger} a + J_{+}J_{-} \rangle_{\Psi} = N \ll M$, we can restrict ourselves to an effective subspace spanned by $\left| M,k \right\rangle $ where $k \ll M$. In this limit, one has $\langle J_z \rangle_{\left| M,k \right\rangle } = -M +2k \approx -M $, that is, $J_z$ acts approximately as an identity and we find $X_3 \approx M a^{\dagger} a - J_{+} J_{-}$. Then, the commutation relations are approximately fulfilled, where $c = \sqrt{M}$. With $\sigma_{\pm} \mathrel{\mathop:}= J_{\pm} / \sqrt{M}$ and $g = \chi \sqrt{M} t$, we find the main result of this section 
\begin{equation}
\label{eq:21}
U \approx e^{-i \tan(g) a \sigma_+} \left( \cos g \right)^{a^{\dagger} a -\sigma_{+}\sigma_{-}}e^{-i \tan(g) a^{\dagger} \sigma_{-}}.
\end{equation}
A straightforward calculation of the action of $U$ on the initial state $| k \rangle \otimes \left| M,0 \right\rangle $ yields 
\begin{equation}
\label{eq:33}
\begin{split}
  & U \left| k \right\rangle \otimes \left| M,0 \right\rangle \approx\\ & 
  \sum_{l=0}^k \sqrt{\binom{k}{l}}\cos^{k-l}(g) (-i \sin g)^l \left|
    k-l \right\rangle \otimes \left| M,l \right\rangle ,
\end{split}
\end{equation} 
which corresponds to a binomial distribution in excitation number basis.
We fix $g = \pi/2$ in order to fully absorb the photonic mode by the spin ensemble. Then, one has that $U \left| k \right\rangle \otimes \left| M,0 \right\rangle  \approx (-i)^k \left| 0 \right\rangle \otimes \left| M,k \right\rangle $. Hence, the physical interpretation of our approximation is that, due to the large number of spins, the probability that a single atom is ``hit by two photons'' is negligible. If we now write the initial photonic state in the Fock basis, $\left| \psi \right\rangle = \sum_k c_k \left| k \right\rangle $ with $c_k \in \mathbbm{C}$, we simply find 
\begin{equation}
\label{eq:22}
U \left| \psi \right\rangle \otimes \left| M,0 \right\rangle \approx \left| 0 \right\rangle \otimes \sum_k  (-i)^k c_k \left| M,k \right\rangle =\mathrel{\mathop:} \left|0  \right\rangle \otimes \left| \phi \right\rangle .
\end{equation}
We denote the set of symmetric states with $\langle J_{+}J_{-} \rangle_{\phi} \leq N \ll M$ by $S_N$.

In the same way as photonic can be mapped to spin states, photonic operators have their spin counterparts. The annihilation operator $a$ is transformed via
\begin{equation}
\label{eq:23}
e^{-i \pi/2 (a \sigma_{+} + a^{\dagger} \sigma_{-})} a e^{i \pi/2 (a \sigma_{+} + a^{\dagger} \sigma_{-})} \approx -\frac{i}{\sqrt{M}} J_{-}.
\end{equation} This can be shown using Eq.~(\ref{eq:20}) with $A = -i g (a \otimes \sigma_{+} +$ $ a^{\dagger} \otimes \sigma_{-}  )$ and $B = a$; noticing that $\left[ A,B \right] = -i \pi/2 \sigma_{-}$ and $\left[ A, \sigma_{-} \right]\approx -i \pi/2 a$.

Finally, let us calculate an interesting example, namely a coherent state $| \alpha \rangle $ as the photonic input state. 
Its amplitudes in the Fock basis read $c_k = \exp(-\left| \alpha \right|^2/2) \alpha^k/\sqrt{k!}$, which is the square root of a Poisson distribution. Note that even though the coherent state contains Fock states $| k \rangle $ that clearly do not satisfy the requirement $k \ll M$, their total contribution is negligible in the limit $M \gg \left| \alpha \right|^2$. We therefore expect only an exponentially small error for the approximation summarized in Eq.~(\ref{eq:21}). Then, the Poisson distribution can be very well approximated by the binomial distribution, that is 
\begin{equation}
\label{eq:24}
c_k \approx \sqrt{\binom{M}{k} }\left(  1-\frac{\left| \alpha \right|^2}{M}\right)^{(M-k)/2} \left( \frac{\alpha}{\sqrt{M}} \right) ^{k},
\end{equation}
if we approximate $(1-\left| \alpha \right|^2/M)^{(M-k)/2} \approx \exp(-\left| \alpha \right|^2/2)$ and $M!/(M-k)! \approx M^k$. Including now the factors $(-i)^k$ as in Eq.~(\ref{eq:22}), we observe that this is exactly the distribution of the product state 
\begin{equation}
\label{eq:26}
\left| \phi_{\alpha} \right\rangle = \left( \sqrt{1- \frac{\left| \alpha \right|^2}{M}}   \left| g \right\rangle -i  \frac{\alpha}{\sqrt{M}}\left| e \right\rangle  \right)^{\otimes M}.
\end{equation}
This is an interesting result, since it shows that the coherent state, which is often considered as the ``most classical'' pure photonic state, does not generate entanglement within the spin ensemble.\footnote{We note that an numerically exact simulation of the interaction (\ref{eq:17}) of a coherent state with the spin ensemble indicates the generation of a small amount of entanglement within the ensemble. We found numerical evidence that, for $g = \pi/2$ and $\left| \alpha \right|^2 \ll M$, the negativity between a fifty-fifty splitting of the spin ensemble equals $\left| \alpha \right|^2/(4M)$.} Instead, every spin (being initially in the ground state) is locally rotated.

In contrast, a pure Fock state $\left| N \right\rangle $ induces much entanglement between the spins, since the respective spin state is the Dicke state $\left| \phi \right\rangle = (-i)^N | M,N \rangle $. Again with the approximation $M \gg N$, a simple expression for the Schmidt decomposition of a $50:50$ splitting between the spins is found to be 
\begin{equation}
\label{eq:27}
\left| M,N \right\rangle = (-i)^N \sum_{l = 0}^N \sqrt{p_l} \left| M/2,l \right\rangle \otimes \left| M/2,N-l \right\rangle ,
\end{equation}
with $p_l   \approx \binom{N}{l}/2^N$. The entanglement between these equally-sized block can then be measured, for example, by the entanglement entropy $E = -\sum_l p_l \log p_l$ \cite{horodecki_quantum_2009}. One finds with further approximations that $E = O(\log N)$.

\section{Applying the photon-spin mapping to measures of macroscopic quantum states}
\label{sec:impl-macro-meas}

In this section, we use the mapping (\ref{eq:22}) to compare measures for photonic and spin systems. Generally, if we use Eq.~(\ref{eq:22}) to map photonic states to spin ensembles, we consider symmetric states from a ``low-energy sector'' $S_N$, which is just a small part of the entire Hilbert space $\mathcal{H}_{\mathrm{S}}$. The maximal scaling of the effective size is supposed to be in the order of the system size of the photonic state, which is the average photon number, $N$. This is a different order than the total number of spins, $M$.

We start by discussing whether the relations (\ref{eq:16}) apply also if we restrict ourselves to $S_N$. In addition, we also study four basic examples of photonic states examined by the ``spin measures'' reviewed in Sec.~\ref{sec:meas-suit-spin} to illustrate our findings. Finally, we discuss direct links between measures for photonic and spin systems.

\subsection{Spin measures used for photonic states: four examples}
\label{sec:spin-measures-used}

The restriction to the subset $S_N$ immediately poses the question whether similar relations like (\ref{eq:16}) hold. Indeed, as shown in \ref{sec:restr-relat-}, one finds that
\begin{equation}
\label{eq:16a}
\begin{split}
  \mathcal{M}^2 =  O(N) &\Leftrightarrow
  \mathcal{N}^{\mathrm{rF}}_{\mathrm{eff}} =  O(N) \Rightarrow \\C_{\delta} =
  O(N) &\Leftrightarrow \bar{D} = O(N) \Rightarrow \\\text{``Modified index } p''
  &\Leftrightarrow \mathcal{N}_{\mathrm{eff}} = O(N)
\end{split}
\end{equation}
for states $| \phi \rangle \in S_N$ (see also Fig.~\ref{fig:connections}). In addition to the assumptions needed for relations (\ref{eq:16}), one has to require the following.

An effective size $C_{\delta} =
O(N) $ implies that the number of spins per measured group is in the order of $M/N \gg 1$. In order to maintain the implication $C_{\delta} =
O(N)  \Rightarrow \mathcal{N}_{\mathrm{eff}} = O(N)$, we have to constrain the measurements to be local (or quasi-local).

Secondly, it is necessary to revise the index $p$. The maximal variance for state from $S_N$ scales with $O(M N)$, that is, no $p = 2$ scaling can be achieved. 
One can still define macroscopic quantum states via the index $p$ for $| \phi \rangle \in S_N$ by dividing the maximal variance by the number of qubits. Then, the state is called macroscopic if this number scales with the maximal scaling, which is here $N$. For pure states, this is mathematically equivalent to the QFI approach (\ref{eq:14}). In the following, we refer to the ``modified index $p$''. Notice that now the quantum fluctuations --for pure states measured by the standard deviation of a local observable-- are not anymore in the order of the spectral radius, which was the initial motivation for the definition of the index $p$. Hence, this slightly modifies the interpretation of the measure \cite{shimizu_stability_2002}. Now, the quantum fluctuations are persistent on a scale which is $N$ times larger than what is expected from ``classical states'' like (spin) coherent states instead of being persistent on the scale linear in $M$.

To illustrate the relations (\ref{eq:16a}), we study four simple examples of photonic states and calculate their effective size for the spin measures of Sec.~\ref{sec:meas-suit-spin}. These are the superposition of two coherent states with opposite phases, $| \Psi_{\alpha} \rangle $, the displaced single photon, $| \psi^{-}_{\alpha} \rangle $, the superposition of two Fock states $| 0 \rangle + \left| 2N \right\rangle $ and a single Fock states $\left| N \right\rangle $. The results are summarized in table \ref{tab:examples}.

The state $| \Psi_{\alpha} \rangle $ translates via the mapping (\ref{eq:22}) into a superposition of two spin-coherent states $\left| \phi_{\alpha} \right\rangle  + \left| \phi_{-\alpha} \right\rangle$ with $\left\langle \phi_{\alpha}| \phi_{-\alpha} \right\rangle = (1-\epsilon^2)^M $ and $\epsilon = 2|\alpha|/ \sqrt{M}\ll 1$ (see Eq.~(\ref{eq:26})).  This state has been intensively investigated in the literature; for instance in Refs.~\cite{dur_effective_2002,bjork_size_2004,korsbakken_measurement-based_2007,marquardt_measuring_2008,frowis_measures_2012}. All these measures agree in assigning an effective size of approximately $M \epsilon^2 = 2 \left| \alpha \right|^2$.

The next example is the displaced single photon state, $| \psi^{-}_{\alpha} \rangle $ (see Eq.~(\ref{eq:37})). It is easy to see that all spin measures are invariant under local rotations. Since the displacement operator is mapped to such a local rotation, the effective size according to these measures is the same as for a Bell state, which is --due to the low number of spin excitations-- necessarily a microscopic state.

The third example is $| 0 \rangle + \left| 2N \right\rangle $ (see Eq.~(\ref{eq:34})), which is mapped to $\left| \phi \right\rangle \propto \left| M,0 \right\rangle + \left|M,2N  \right\rangle$.  The interference-based measure \cite{bjork_size_2004} has some problems dealing with this state. With the original definition (\ref{eq:10}), it diverges for the choice $A = a^{\dagger} a $, which correspond to $ J_{+}J_{-}/M$ in the spin case. If we instead maximize the denominator of $\mathcal{M}^2$ as suggested in Ref.~\cite{frowis_measures_2012}, the measure becomes vanishingly small, since $\mathcal{V}_{\left| 2N \right\rangle } (J_x) \approx M (4N + 1)$ while the difference in the expectation value of $a^{\dagger} a$ scales with $N$.

The relative Fisher information \cite{frowis_measures_2012} performs similarly. The reason is that the maximal Fisher information is already large for high-excitation Dicke states. We find that $\mathcal{N}_{\mathrm{eff}}(\phi) = 2N + 1$, while one has $\mathcal{N}_{\mathrm{eff}}(\left| M,0 \right\rangle ) = 1$ and $\mathcal{N}_{\mathrm{eff}}(\left| M,2N \right\rangle ) = 4N + 1$. With Eq.~(\ref{eq:15}), one has $\mathcal{N}_{\mathrm{eff}}^{\mathrm{rF}}(\phi) = 1$. The relative Fisher information does not identify the state of Eq.~(\ref{eq:34}) as a Schr\"odinger-cat state since already one constituent is considered to be a macroscopic quantum state.

The measurement-based measure \cite{korsbakken_measurement-based_2007}, on the other side, assigns to this state a large effective size. To see this, divide the spin ensemble into $x$ groups. The reduced density matrices of one such group read  $\rho_0^x = \left| M/x,0 \right\rangle\!\left\langle M/x,0\right| $ and $\rho_{2N}^{(M/x)} = \sum_{k = 0}^{2N}  p_k^{(M/x)}  \left| M/x,k \right\rangle \left\langle M/x,k\right| $, with
\begin{equation}
\label{eq:28}
p_k^{(M/x)} \approx \binom{2N}{k}(1/x)^k (1-1/x)^{2N-k},
\end{equation}
respectively. Therefore, the success probability (\ref{eq:11}) to distinguish the two states equals $P_{\mathrm{S}}^{(M/x)} = 1-p_0^x/2 \approx 1- (1-1/x)^{2N}/2$. In order to distinguish with sufficiently high probability for any $N$, one can maximally set $x = c N$ with proper value of $c$ depending on the given threshold $\delta$. Hence, we find $C_{\delta} = O(N)$.

Similarly, the ``microscopic-step''-based measure \cite{marquardt_measuring_2008} assigns a large effective size to the state (\ref{eq:34}), since one needs exactly $2N$ step in ``units'' of $J_{+}$ in order to reach $\left| 2N \right\rangle $ from the vacuum.

The QFI-based measure and the (modified) index $p$ recognizes this state as macroscopic, since, as mentioned before, Dicke states themselves are macroscopic quantum states provided $N$ is large enough.

The last example, the single Fock state $| N \rangle $ can only be treated by the Fisher information and the index $p$. Again, one finds that $\mathcal{N}_{\mathrm{eff}} \approx 2 N +1$.

\begin{table}[htbp]
  \begin{tabular}{l  l l l l l l }  \hline  \hline  
    Eff. size
    &  Eq. & $| \Psi_{\alpha} \rangle $ & $| \psi^{-}_{\alpha}  \rangle $ & $| 0 \rangle + \left| 2N \right\rangle $ & $| N \rangle $\\
    \hline 
    $\mathcal{M}^2$ 
    & (\ref{eq:10}) & $O(N)$   & $O(1)$& $O(1/M)$ & n.d. \\
    $N_{\mathrm{eff}}^{\mathrm{rF}}$ 
    & (\ref{eq:15})
    & 
$O(N)$    & $O(1)$& $O(1)$ & n.d. \\ 
$C_{\delta}$ 
& (\ref{eq:12}) & $O(N)$  & $O(1)$& $O(N)$  & n.d. \\
$\bar{D}$ 
& (\ref{eq:13}) & $O(N)$  & $O(1)$& $O(N)$  & n.d. \\
Size$_{P_g}$ 
& (\ref{eq:25}) & $O(N)$  & $O(\sqrt{N})$& $O(N)$  & n.d. \\
Index $p^{(*)}$ 
& (\ref{eq:7}) &  $O(N)$  & $O(1)$& $O(N)$  & $O(N)$ \\
$N_{\mathrm{eff}}$
& (\ref{eq:14})& $O(N)$  & $O(1)$& $O(N)$  & $O(N)$ \\
$\mathcal{I}$ 
& (\ref{eq:4}) & $O(N)$  & $O(1)$& $O(N)$  & $O(N)$  \\
  \hline
\end{tabular}
  \caption{Four examples of photonic states and their effective sizes according to the discussed measures. The second column guides the reader to the definitions of the measures. The remaining columns summarize the effective sizes for the four examples discussed in Sec.~\ref{sec:spin-measures-used}. ``n.d.'' stands for ``not defined''. $^{(*)}$ The index $p$ is used in the modified version (see text).{\label{tab:examples}}}
\end{table}

\subsection{Measurement-based vs.~coarse-grained-based measures}
\label{sec:meas-based-vs}

Let us now come to a direct comparison between the measurement-based measure for spin system \cite{korsbakken_measurement-based_2007}, $C_{\delta}$, and the coarse-grained-based measure formulated for photonic modes \cite{sekatski_size_2014}, $\mathrm{Size}_{P_g}$. The similarity between the measures is clearly the motivation to call two states $| \phi_0 \rangle $ and $| \phi_1 \rangle $ macroscopically distinct if they stay distinguishable under special restrictions. In the former case, it is the limited access to only a subset of all particles. The latter select insensitive photon number detectors as a realistic constraint in high-photon-number scenarios.

As we have seen in Eq.~(\ref{eq:28}), the reduced density matrix of Dicke state is a binomially distributed, incoherent mixture of Dicke states. If we translate the Dicke states back to photonic Fock states, we see that this is equivalent of considering the Fock state after a beam splitter with transmitivity $1/x$, where the reflected part is traced out. The photon number detection right after the beam splitter therefore corresponds to an \textit{inefficient} detector. It is an open question how far reaching the similarities of the two measures are. In particular, does $C_{\delta} = O(N)$ imply that $\mathrm{Size}_{P_g} = O(N)$ and vice-versa? Even if this is not the case, there is clearly a strong connection between the measures (and the microscopic-step-based measure \cite{marquardt_measuring_2008}). We thus conclude that the two measures are equivalent up to the choice of the noise model for the detection. It is clear that  this choice depends on the experimental situation and, for theoretical studies, one may even consider both measures in parallel.

\subsection{Comparing measures for general macroscopic quantum states}
\label{sec:comp-meas-gener}

Although the Wigner-function-based measure \cite{lee_quantification_2011}, the index $p$ \cite{shimizu_stability_2002} and the QFI-based measure \cite{frowis_measures_2012} are motivated by different arguments, we will see in this section that there exist tight mathematical connections between them.

To this end, we map the entire measure \cite{lee_quantification_2011} to the spin ensemble via $a \mapsto -i/\sqrt{M} J_{-}$ (see Eq.~(\ref{eq:23})). With the identity $J_{\pm} = 1/2 (J_{x} \pm i J_{-})$, it is straightforward to see that 
\begin{equation}
\label{eq:29}
\mathcal{I}(\rho) \mapsto \frac{1}{8M} \left( \langle  \left[ J_x,[J_x, \rho] \right]\rangle_{\rho} +  \langle\left[ J_y,[J_y, \rho] \right]\rangle_{\rho}\right),
\end{equation}
which reduced in the case of pure states to 
\begin{equation}
\label{eq:30}
\mathcal{I}(\psi) \mapsto \frac{1}{4M} \left( \mathcal{V}_{\psi}(J_x) + \mathcal{V}_{\psi}(J_y) \right).
\end{equation}
We compare this to the QFI-based measure (\ref{eq:14}) for pure states, which is the maximal variance for local observables divided by $M$. The local operator leading to the maximal variance must be located on the equator, that is, $J_{\mathrm{max}} = \cos \varphi J_x + \sin \varphi J_y$ for a specific value of $\varphi$, because the contribution from $J_z$ is at most in the order of $N^2$, which is significantly lower than the maximal variance, which can be in the order of $MN$. Thus, it is straightforward to bound $\mathcal{N}_{\mathrm{eff}}(\phi)/4 < \mathcal{I}(\phi) \leq  \mathcal{N}_{\mathrm{eff}}(\phi)/2$.
We find the same result for the index $p$, given the modification of the effective size discussed in Sec.~\ref{sec:spin-measures-used}.

In contrast to the measures for Schr\"odinger-cat states such as Eq.~(\ref{eq:1}), the measures discussed in this section identify Fock states (Dicke states)  and (spin) squeezed states as macroscopic quantum states, if the mean photon number, $\langle a^{\dagger}a \rangle$, is high enough. The effective size of Fock states $\left| N \right\rangle $ and squeezed vacuum states is $\langle a^{\dagger}a \rangle + 1/2$ \cite{lee_quantification_2011}, which is the maximal number.

In the case of mixed states, we still find some connection between the measures. First note that $\mathcal{I}(\rho)$ is a convex function, that is, for $\rho = a \rho_1 + (1-a)\rho_2$, one has that $\mathcal{I}(\rho) = a \mathcal{I}(\rho_1) + (1-a) \mathcal{I}(\rho_2) - a(1-a) \mathcal{I}(\rho_1-\rho_2)$. Next, it was conjectured \cite{toth_extremal_2013} and later proven \cite{yu_quantum_2013} that the QFI is the convex roof of the variance (up to a factor of four). This implies that any convex function that reduces to the variance for pure states can be bounded from above by the QFI. Thus, one has 
\begin{equation}
\label{eq:31}
\mathcal{I}(\rho) \leq \frac{1}{2} \mathcal{N}_{\mathrm{eff}}(\rho).
\end{equation}
Furthermore, the trace norm in the definition of index $q$ (see Eq.~(\ref{eq:8})) is also an upper bound on $\mathcal{I}(\rho)$, since for any operator $A,B$ it holds that $\mathrm{Tr}(A B) \leq \lVert B \rVert \lVert A \rVert_1$; in our case, $B = \rho$ with operator norm $\lVert \rho \rVert$ smaller or equal unity.

Even though the full equivalence between the measures for mixed states is an open question, they agree on the effective size for the following state 
\begin{equation}
\label{eq:32}
\rho = \frac{1 + d}{2} \left| \Psi_{\alpha} \right\rangle\!\left\langle \Psi_{\alpha}\right| + \frac{1 - d}{2} \left| \Psi_{\alpha}^{-} \right\rangle\!\left\langle \Psi_{\alpha}^{-}\right|,
\end{equation}
where $| \Psi_{\alpha}^{-} \rangle \propto \left| \alpha \right\rangle - \left| -\alpha \right\rangle $ and $d \in [0,1]$. All three measures assign a value in the order of $d^2 \left| \alpha \right|^2$ to this quantum state (for $\mathcal{I}(\rho)$, see Ref.~\cite{lee_quantification_2011}).

\begin{figure}[htbp]
\centerline{\includegraphics[width=\columnwidth]{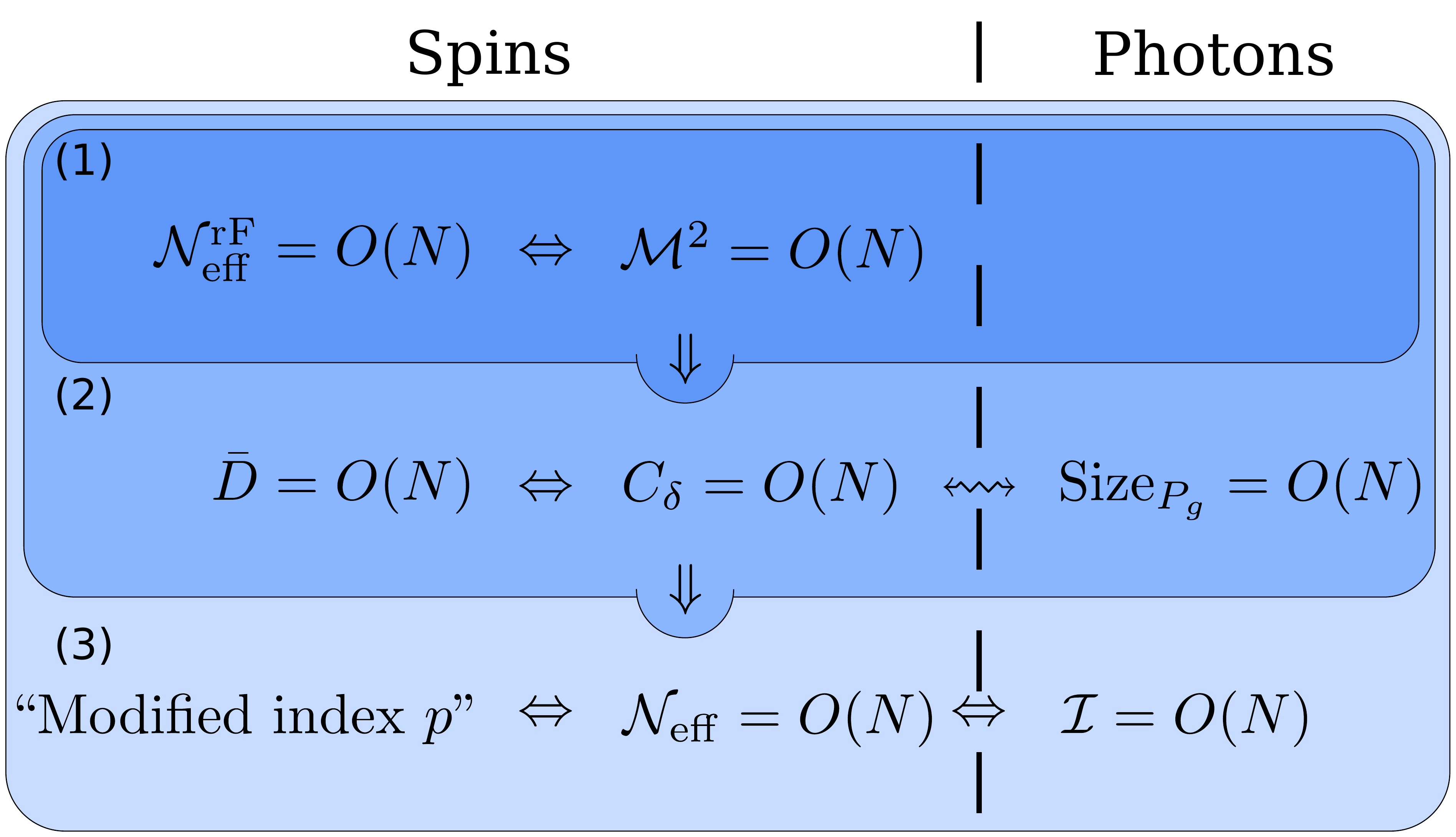}}
\caption[]{\label{fig:connections} Summary of the relations between the measures for states $\left| \phi \right\rangle \in S_N$ as discussed in Sec.~\ref{sec:impl-macro-meas}. We identify three different levels of macroscopic quantum states, where each level includes the previous one: (1) superpositions of ``classical'' states like $| \Psi_{\alpha} \rangle $; (2) superpositions of ``macroscopically distinct'' states like $\left| 0 \right\rangle + \left| 2N \right\rangle $; (3) general macroscopic quantum states like $| N \rangle $. The implications are either mathematically strict ($\Rightarrow, \Leftrightarrow$) or there is a conceptual connection ($\leftrightsquigarrow
$). The assumptions and restrictions that are used for the relations are discussed in the text. Compare also to table \ref{tab:examples}.}
\end{figure}

The results of this section are summarized in Fig.~\ref{fig:connections}.

\section{Summary and discussion}
\label{sec:discussion}

In this paper, we considered an idealized photon-spin interaction to fully map one-mode photonic states onto spin ensembles in a regime where the spin number, $M$,  is much larger than the mean photon number, $N$. Thus we have a simple tool at hand to compare photonic states with spin states. More specifically, we are able to contrast measures for macroscopic quantum states for photonic and spin systems. To exemplify the usefulness of the mapping, we presented three main results.

First, we showed that measures for spin systems essentially keep the relations among them even if we restrict to spin states that result from the absorption of the photonic excitations. It is important to stress that operating in the limit $N \ll M$ requires a ``redefinition'' of the notion of ``macroscopic quantum state''. The effective size of these states can never go beyond $O(N)$, while quantum states from the full Hilbert space may exhibit effective sizes scaling as $O(M)$. 

Second, we found similarities between the coarse-grained-based measure for photonic modes \cite{sekatski_size_2014} and the measurement-based measure mainly for spin ensembles \cite{korsbakken_measurement-based_2007} as both measures directly or indirectly connect the notion of macroscopic distinctness to distinctness under macroscopically realistic conditions. They differ in the exact modeling of this idea, but they are tightly related on a conceptual level and also lead to comparable results for the discussed examples.

Finally, we identified strong mathematical connections between measures for general macroscopic quantum states \cite{shimizu_stability_2002,shimizu_detection_2005,lee_quantification_2011,frowis_measures_2012}. This is interesting as the motivation for these measures is partially very different and nevertheless they give rise to a similar classification of macroscopic quantum states. 

One may ask how unambiguous this mapping is and whether, for example, more realistic models could lead to different conclusions. It is clear that there exists indeed such mappings; for instance, if every Fock state is not mapped to a coherent but to an incoherent mixture of excitations. Then, the photonic state does not induce entanglement within the spin ensemble and the measures do not assign any large effective size. However, this choice for the mapping does not preserve the properties of the photonic state since it is not reversible. We further note that similar results as presented in this paper can be obtained by more realistic Hamiltonians. For example, one may respect some geometrical aspect like the spatial distribution of the spins and the direction of the incoming light beam. As long as the photons potentially generate sufficiently large entanglement between the spins, all conclusions of this paper are still valid.
Even more, we hope that the presented mapping is useful for other theoretical studies to characterize the nonclassical properties of some photonic states, since notions like entanglement or nonlocality are  not directly defined for a single-mode states (cp., e.g., to Refs.~\cite{asboth_computable_2005,ivan_generation_2006}).

Through this mapping, we have discovered strong connections between the discussed measures. However, one clearly encounters conceptual differences. For example, the coarse-grained-based measure \cite{sekatski_size_2014} for single modes is not invariant under displacement, while  all other mentioned measures are. In particular, for spin ensembles, the displacement operator translates to a local rotation of the spins. Even though this seems to be a rather small detail, it is important to clarify such open questions. Ultimately, it is desirable to develop a ``resource theory for macroscopic quantumness'', similar as it was done for entanglement \cite{skotiniotis_2014}. This theory would be able to answer questions like: Under which unitary operations is the effective invariant (like local rotations for entangled states)? It is a worthwhile observation that, for spin systems, one can increase the effective size by local measurements. For example, a two-dimensional cluster state \cite{briegel_persistent_2001} has a small variance with respect to any (quasi-)local operator and therefore a small effective size. However, by measuring every second qubit, one can deterministically generate a GHZ state (up to known local unitaries) with an effective size equaling $M/2$.

It is further an interesting question whether this approach of comparing measures for macroscopic quantum states can be extended to compare, for instance, massive bosonic modes with light modes. In this case, archetypal optomechanical setups could be used for a mapping between light and massive objects. However, there are also limitations and it is likely that even on a qualitative level there could remain certain differences between proposals for macroscopic quantum states. In Ref.~\cite{nimmrichter_macroscopicity_2013}, for example, the notion of macroscopic quantum states is linked to states whose generation is potentially used for a falsification of collapse models like \cite{ghirardi_unified_1986}. However, the influence of the mass of the object onto measures for photonic systems is an open question and thus a direct comparison is nontrivial.

Altogether, this work shows that the discussed measures share common features. In particular, there appears a hierarchy of three different levels of macroscopic quantum states. There are superpositions of ``classical'' states, superpositions of ``macroscopically distinct'' states and general macroscopic quantum states that show some sort of macroscopic quantum effect. We also see that it is sometimes not straightforward to ``transfer'' the measures to other physical systems. It may happen that the resulting effective sizes are to some extent surprising, as for example states like $| 0 \rangle + \left| 2N \right\rangle $ are not considered to be macroscopically quantum by some measures. The restriction to the low-energy sector reveals conflicts of some measures with our intuition. Although this does not imply that these measures are not well-defined, it may show us that measures based on some intuition are mainly applicable to a regime where the intuition comes from. However, the usage or adaption of measures to diversity of different scenarios and experimental setups helps to understand the features as well as the limitations of each approach. Eventually, one may wish to modify the proposals to be able to handle a wider range of experiments. We therefore hope that this work contribute to a better understanding of the structure of the different measures and it might be one step towards a commonly accepted theory of macroscopic quantum physics.

\textit{Acknowledgments.--- } We thank Pavel Sekatski for helpful discussions and comments. This work was supported by the Austrian Science Fund (FWF), grant number J3462; the National Swiss Science Foundation (SNSF), grant number PP00P2-150579; and the European Research Council (MEC).

\appendix

\section{Restrictions for relations (\ref{eq:16}) and (\ref{eq:16a})}
\label{sec:restr-relat-}

\textit{Sufficient restrictions for relations (\ref{eq:16}).--- } As discussed in Ref.~\cite{frowis_measures_2012}, the following assumptions are proceeded on to show the relations of Eq.~(\ref{eq:16}). For $\mathcal{M}^2$, we maximize numerator and denominator independently over all local operators in order maintain finite effective sizes. For $C_{\delta}$, we assume that the scaling of the effective size $C_{\delta}$ does not depend on the specific choice of $\delta$. A sufficient condition therefore is to demand that there are negligible correlations in the measurement statistics between the subgroups.  For $\bar{D}$, we restrict ourselves to states where $\bar{D}$ is independent whether one counts the steps from $|\psi_0  \rangle $ to $| \psi_1 \rangle $ or vice-versa.

\textit{Additional arguments for relations (\ref{eq:16a}).--- } To show the relations (\ref{eq:16a}), it suffices to slightly modify the proof for (\ref{eq:16}) in Ref.~\cite{frowis_measures_2012}. First, we show that for states $| \phi \rangle \in S_N$, the maximal variance is bounded by $O(MN)$. To see this, notice that symmetric states are eigenstates of the total angular momentum, $J^{2} = J_x^2 + J_y^2 + J_z^2$, with eigenvalue $M(M + 2)$. Since $\langle J_z \rangle_{\phi} = -M + 2N$, we find that $\max_{J: \mathrm{local}} V_{\phi}(J) \leq \mathcal{V}_{\phi}(J_x) + \mathcal{V}_{\phi}(J_t) + \mathcal{V}_{\phi}(J_z) \leq \langle J^2 \rangle_{\phi} - \langle J_z \rangle_{\phi}^2 \approx 4M (N + 1/2)$. So one finds 
\begin{equation}
\label{eq:35}
\max_{\phi \in S_N} \max_{J: \mathrm{local}} V_{\phi}(J)= O(NM).
\end{equation}
This observation also motivates the modification of the index $p$ for states from $S_N$.

Similarly, one can show that the maximal expectation value for any local operator of the form $J_{\vartheta} = \cos \vartheta J_x + \sin \vartheta J_y$ is in the order of $\sqrt{MN}$. Consider the reduced density matrix for one spin, which equals $\rho_1 = 1/2( \mathbbm{1} + \vec{n} \vec{\sigma})$. For states $\left| \phi \right\rangle \in S_N$, we thus have $\mathrm{Tr(\rho_1 \sigma_z)} \leq -1+ 2 N/M$. Hence, one finds that $n_z= 1 - 2 N/M$ and, due to the positivity of $\rho_1$, that $n_x^2 + n_y^2 \leq 4N/M$. So, we can estimate $\max_{\vartheta} \langle J_{\vartheta} \rangle_{\phi} \leq M \sqrt{4N/M} = O(\sqrt{M N})$. Furthermore, this shows that $\max_{\left| \phi_0 \right\rangle,\left| \phi_1 \right\rangle  \in S_N} \left| \langle J_z \rangle_{\phi_0} - \langle J_z \rangle_{\phi_1} \right| = O(N)$. 

The estimates on the maximal variance and the maximal expectation values are important to bound the effective sizes of $\mathcal{M}^2$, $\mathcal{N}_{\mathrm{eff}}^{\mathrm{rF}}$ and $\mathcal{N}_{\mathrm{eff}}$. With these findings, one argues that the effective size of these measures are at most $O(N)$.

In addition, we have to consider the following issue.
For every state family where $C_{\delta}$ does not scale linearly in $M$, it is clear that the minimal group size $n_{\min}$ depends on $M$. In particular, for $C_{\delta} = O(N)$, one has that $n_{\min} = O(M/N)$. If we allow for general measurements on such large groups, we cannot establish a direct link to the QFI approach $\mathcal{N}_{\mathrm{eff}}$, because a nonlocal measurement is translated to a nonlocal Hamiltonian, which is not the class that is considered for the maximization of $\mathcal{V}(H)$. Therefore, in order to maintain this relation, we have to restrict ourselves to collective measurements.

 %%%%%%%%%%%%%%%%%%%%%%%%%%%%%%%%%%%%%%%%%%%%%%%%%%%%%%%%%%%%%%%%%%%%%%%%%%%%%%%%%%%%%%

%\section*{References}
%\label{sec:references}

\bibliographystyle{elsarticle-num}

\bibliography{References}

\begin{thebibliography}{10}
\expandafter\ifx\csname url\endcsname\relax
  \def\url#1{\texttt{#1}}\fi
\expandafter\ifx\csname urlprefix\endcsname\relax\def\urlprefix{URL }\fi
\expandafter\ifx\csname href\endcsname\relax
  \def\href#1#2{#2} \def\path#1{#1}\fi

\bibitem{andrews_observation_1997}
M.~R. Andrews, C.~G. Townsend, H.-J. Miesner, D.~S. Durfee, D.~M. Kurn,
  W.~Ketterle, Observation of interference between two bose condensates,
  Science 275~(5300) (1997) 637--641.

\bibitem{friedman_quantum_2000}
J.~R. Friedman, V.~Patel, W.~Chen, S.~K. Tolpygo, J.~E. Lukens, Quantum
  superposition of distinct macroscopic states, Nature 406~(6791) (2000)
  43--46.

\bibitem{hime_solid-state_2006}
T.~Hime, P.~A. Reichardt, B.~L.~T. Plourde, T.~L. Robertson, C.-E. Wu, A.~V.
  Ustinov, J.~Clarke, Solid-state qubits with current-controlled coupling,
  Science 314~(5804) (2006) 1427--1429.

\bibitem{teufel_sideband_2011}
J.~D. Teufel, T.~Donner, D.~Li, J.~W. Harlow, M.~S. Allman, K.~Cicak, A.~J.
  Sirois, J.~D. Whittaker, K.~W. Lehnert, R.~W. Simmonds, Sideband cooling of
  micromechanical motion to the quantum ground state, Nature 475~(7356) (2011)
  359--363.

\bibitem{kiesel_cavity_2013}
N.~Kiesel, F.~Blaser, U.~Deli{\'c}, D.~Grass, R.~Kaltenbaek, M.~Aspelmeyer,
  Cavity cooling of an optically levitated submicron particle, {PNAS} (2013)
  201309167.

\bibitem{gerlich_quantum_2011}
S.~Gerlich, S.~Eibenberger, M.~Tomandl, S.~Nimmrichter, K.~Hornberger, P.~J.
  Fagan, J.~T{\"u}xen, M.~Mayor, M.~Arndt, Quantum interference of large
  organic molecules, Nat Commun 2 (2011) 263.

\bibitem{lee_entangling_2011}
K.~C. Lee, M.~R. Sprague, B.~J. Sussman, J.~Nunn, N.~K. Langford, X.-M. Jin,
  T.~Champion, P.~Michelberger, K.~F. Reim, D.~England, D.~Jaksch, I.~A.
  Walmsley, Entangling macroscopic diamonds at room temperature, Science
  334~(6060) (2011) 1253--1256.

\bibitem{hald_spin_1999}
J.~Hald, J.~L. S{\o}rensen, C.~Schori, E.~S. Polzik, Spin squeezed atoms: A
  macroscopic entangled ensemble created by light, Phys. Rev. Lett. 83~(7)
  (1999) 1319--1322.

\bibitem{usmani_heralded_2012}
I.~Usmani, C.~Clausen, F.~Bussi{\`e}res, N.~Sangouard, M.~Afzelius, N.~Gisin,
  Heralded quantum entanglement between two crystals, Nature Photonics 6~(4)
  (2012) 234--237.

\bibitem{riedel_atom-chip-based_2010}
M.~F. Riedel, P.~B{\"o}hi, Y.~Li, T.~W. H{\"a}nsch, A.~Sinatra, P.~Treutlein,
  Atom-chip-based generation of entanglement for quantum metrology, Nature
  464~(7292) (2010) 1170--1173.

\bibitem{behbood_generation_2014}
N.~Behbood, F.~M. Ciurana, G.~Colangelo, M.~Napolitano, G.~Toth, R.~J. Sewell,
  M.~W. Mitchell, Generation of macroscopic singlet states in a cold atomic
  ensemble, {arXiv:1403.1964} [quant-ph].

\bibitem{lucke_detecting_2014}
B.~L{\"u}cke, J.~Peise, G.~Vitagliano, J.~Arlt, L.~Santos, G.~T{\'o}th,
  C.~Klempt, Detecting multiparticle entanglement of dicke states,
  {arXiv:1403.4542} [cond-mat, physics:quant-ph].

\bibitem{bruno_displacement_2013}
N.~Bruno, A.~Martin, P.~Sekatski, N.~Sangouard, R.~T. Thew, N.~Gisin,
  Displacement of entanglement back and forth between the micro and macro
  domains, Nat Phys 9~(9) (2013) 545--548.

\bibitem{lvovsky_observation_2013}
A.~I. Lvovsky, R.~Ghobadi, A.~Chandra, A.~S. Prasad, C.~Simon, Observation of
  micro-macro entanglement of light, Nat Phys 9~(9) (2013) 541--544.

\bibitem{zurek_decoherence_2003}
W.~H. Zurek, Decoherence, einselection, and the quantum origins of the
  classical, Rev. Mod. Phys. 75~(3) (2003) 715--775.

\bibitem{ghirardi_unified_1986}
G.~C. Ghirardi, A.~Rimini, T.~Weber, Unified dynamics for microscopic and
  macroscopic systems, Phys. Rev. D 34~(2) (1986) 470--491.

\bibitem{bassi_models_2013}
A.~Bassi, K.~Lochan, S.~Satin, T.~P. Singh, H.~Ulbricht, Models of
  wave-function collapse, underlying theories, and experimental tests, Rev.
  Mod. Phys. 85~(2) (2013) 471--527.

\bibitem{schrodinger_gegenwartige_1935}
E.~Schr{\"o}dinger, Die gegenw{\"a}rtige situation in der quantenmechanik,
  Naturwissenschaften 23~(48) (1935) 807--812.

\bibitem{leggett_macroscopic_1980}
A.~J. Leggett, Macroscopic quantum systems and the quantum theory of
  measurement, Progress of Theoretical Physics Supplement 69 (1980) 80--100.

\bibitem{leggett_testing_2002}
A.~J. Leggett, Testing the limits of quantum mechanics: motivation, state of
  play, prospects, Journal of Physics: Condensed Matter 14~(15) (2002)
  R415--R451.

\bibitem{dur_effective_2002}
W.~D{\"u}r, C.~Simon, J.~I. Cirac, Effective size of certain macroscopic
  quantum superpositions, Phys. Rev. Lett. 89~(21) (2002) 210402.

\bibitem{shimizu_stability_2002}
A.~Shimizu, T.~Miyadera, Stability of quantum states of finite macroscopic
  systems against classical noises, perturbations from environments, and local
  measurements, Phys. Rev. Lett. 89~(27) (2002) 270403.

\bibitem{bjork_size_2004}
G.~Bj{\"o}rk, P.~G.~L. Mana, A size criterion for macroscopic superposition
  states, J. Opt. B: Quantum Semiclass. Opt. 6~(11) (2004) 429.

\bibitem{shimizu_detection_2005}
A.~Shimizu, T.~Morimae, Detection of macroscopic entanglement by correlation of
  local observables, Phys. Rev. Lett. 95~(9) (2005) 090401.

\bibitem{korsbakken_measurement-based_2007}
J.~I. Korsbakken, K.~B. Whaley, J.~Dubois, J.~I. Cirac, Measurement-based
  measure of the size of macroscopic quantum superpositions, Phys. Rev. A
  75~(4) (2007) 042106.

\bibitem{marquardt_measuring_2008}
F.~Marquardt, B.~Abel, J.~von Delft, Measuring the size of a quantum
  superposition of many-body states, Phys. Rev. A 78~(1) (2008) 012109.

\bibitem{lee_quantification_2011}
C.-W. Lee, H.~Jeong, Quantification of macroscopic quantum superpositions
  within phase space, Phys. Rev. Lett. 106~(22) (2011) 220401.

\bibitem{frowis_measures_2012}
F.~Fr{\"o}wis, W.~D{\"u}r, Measures of macroscopicity for quantum spin systems,
  New Journal of Physics 14~(9) (2012) 093039.

\bibitem{nimmrichter_macroscopicity_2013}
S.~Nimmrichter, K.~Hornberger, Macroscopicity of mechanical quantum
  superposition states, Phys. Rev. Lett. 110~(16) (2013) 160403.

\bibitem{sekatski_size_2014}
P.~Sekatski, N.~Sangouard, N.~Gisin, Size of quantum superpositions as measured
  with classical detectors, Phys. Rev. A 89~(1) (2014) 012116.

\bibitem{frowis_certifiability_2013}
F.~Fr{\"o}wis, M.~v.~d. Nest, W.~D{\"u}r, Certifiability criterion for
  large-scale quantum systems, New J. Phys. 15~(11) (2013) 113011.

\bibitem{sekatski_how_2014}
P.~Sekatski, N.~Gisin, N.~Sangouard, How difficult it is to prove the
  quantumness of macroscropic states?, {arXiv:1402.2542} [quant-ph].

\bibitem{volkoff_measurement-_2014}
T.~J. Volkoff, K.~B. Whaley, Measurement- and comparison-based sizes of
  schr{\"o}dinger cat states of light, Phys. Rev. A 89~(1) (2014) 012122.

\bibitem{sekatski_proposal_2012}
P.~Sekatski, N.~Sangouard, M.~Stobi{\'n}ska, F.~Bussi{\`e}res, M.~Afzelius,
  N.~Gisin, Proposal for exploring macroscopic entanglement with a single
  photon and coherent states, Phys. Rev. A 86~(6) (2012) 060301.

\bibitem{jeong_reply_2011}
H.~Jeong, M.~Kang, C.-W. Lee, Reply to comment on {"Quantification} of
  macroscopic quantum superpositions within phase space", {arXiv:1108.0212}
  [quant-ph].

\bibitem{morimae_superposition_2010}
T.~Morimae, Superposition of macroscopically distinct states means large
  multipartite entanglement, Phys. Rev. A 81~(1) (2010) 010101.

\bibitem{mandelstam_uncertainty_1945}
L.~Mandelstam, I.~Tamm, The uncertainty relation between energy and time in
  non-relativistic quantum mechanics, J. Phys. ({UDSSR)} 9~(4) (1945) 249.

\bibitem{braunstein_statistical_1994}
S.~L. Braunstein, C.~M. Caves, Statistical distance and the geometry of quantum
  states, Phys. Rev. Lett. 72~(22) (1994) 3439.

\bibitem{frowis_kind_2012}
F.~Fr{\"o}wis, Kind of entanglement that speeds up quantum evolution, Phys.
  Rev. A 85~(5) (2012) 052127.

\bibitem{toth_multipartite_2012}
G.~T{\'o}th, Multipartite entanglement and high-precision metrology, Phys. Rev.
  A 85~(2) (2012) 022322.

\bibitem{hyllus_fisher_2012}
P.~Hyllus, W.~Laskowski, R.~Krischek, C.~Schwemmer, W.~Wieczorek,
  H.~Weinfurter, L.~Pezz{\'e}, A.~Smerzi, Fisher information and multiparticle
  entanglement, Phys. Rev. A 85~(2) (2012) 022321.

\bibitem{frowis_are_2012}
F.~Fr{\"o}wis, W.~D{\"u}r, Are cloned quantum states macroscopic?, Phys. Rev.
  Lett. 109~(17) (2012) 170401.

\bibitem{hammerer_quantum_2010}
K.~Hammerer, A.~S. S{\o}rensen, E.~S. Polzik, Quantum interface between light
  and atomic ensembles, Rev. Mod. Phys. 82~(2) (2010) 1041--1093.

\bibitem{horodecki_quantum_2009}
R.~Horodecki, P.~Horodecki, M.~Horodecki, K.~Horodecki, Quantum entanglement,
  Rev. Mod. Phys. 81~(2) (2009) 865--942.

\bibitem{toth_extremal_2013}
G.~T{\'o}th, D.~Petz, Extremal properties of the variance and the quantum
  fisher information, Phys. Rev. A 87~(3) (2013) 032324.

\bibitem{yu_quantum_2013}
S.~Yu, Quantum fisher information as the convex roof of variance,
  {arXiv:1302.5311}.

\bibitem{asboth_computable_2005}
J.~K. Asb{\'o}th, J.~Calsamiglia, H.~Ritsch, Computable measure of
  nonclassicality for light, Phys. Rev. Lett. 94~(17) (2005) 173602.

\bibitem{ivan_generation_2006}
J.~S. Ivan, N.~Mukunda, R.~Simon, Generation of {NPT} entanglement from
  nonclassical photon statistics, {arXiv:quant-ph/0603255}.

\bibitem{skotiniotis_2014}
M.~Skotiniotis, Private communication, 2014.

\bibitem{briegel_persistent_2001}
H.~J. Briegel, R.~Raussendorf, Persistent entanglement in arrays of interacting
  particles, Phys. Rev. Lett. 86~(5) (2001) 910--913.

\end{thebibliography}

\end{document}